\newcommand{\be}{\begin{equation}}
\newcommand{\ee}{\end{equation}}
\newcommand{\lb}{\label}
\newcommand{\br}{{\bf r}}
\newcommand{\bv}{{\bf v}}
\newcommand{\bM}{{\bf M}}
\newcommand{\bR}{{\bf R}}
\newcommand{\bP}{{\bf P}}
\newcommand{\bG}{{\bf G}}
\newcommand{\bV}{{\bf V}}
\newcommand{\bz}{{\bf 0}}
\newcommand{\oH}{\widehat{{\cal H}}}
\newcommand{\cL}{{\cal L}}
\newcommand{\grad}{{\mbox{\boldmath $\nabla$}}}
\newcommand{\bdot}{{\mbox{\boldmath $\cdot$}}}
\documentstyle[twocolumn,prl,aps]{revtex}
\begin{document}
\relax
\draft \preprint{}
\title{Universality of the Inertial-Convective Range in Kraichnan's Model
       of a Passive Scalar}
\author{Gregory Eyink and Jack Xin}
\address{Department of Mathematics\\
         University of Arizona}
\date{\today}
\maketitle
\begin{abstract}
We establish by exact, nonperturbative methods a universality for the
correlation
functions in Kraichnan's ``rapid-change'' model of a passively advected scalar
field. We
show that the solutions for separated points in the convective range of scales
are unique
and independent of the particular mechanism of the scalar dissipation. Any
non-universal
dependences therefore must arise from the large length-scale features. The main
step in the proof is to show
that solutions of the model equations are unique even in the idealized case of
zero diffusivity, under
a very modest regularity requirement (square-integrability). Within this
regularity class the only
zero-modes of the global many-body operators are shown to be trivial ones (i.e.
constants). In a bounded
domain of size $L$, with physical boundary conditions, the ``ground-state
energy'' is strictly positive
and scales as $L^{-\gamma}$ with an exponent $\gamma >0$.
\end{abstract}

\medskip

\medskip


\narrowtext
\medskip

\medskip

\medskip

\medskip

The Kolmogorov 1941 theory (K41) \cite{K41} of fully developed turbulence
hypothesized the existence
of universal statistics at so-called {\em inertial-range} scales $L\gg
\ell\gg\eta,$ where $L$ is the {\em integral
length} characteristic of the peak energy and $\eta$ is the {\em (Kolmogorov)
microscale} characteristic of
the peak dissipation. According to the first and second similarity hypotheses
of Kolmogorov \cite{K41},
both the limits $L\rightarrow \infty$ and $\eta\rightarrow 0$ should exist for
the distribution function
(PDF) of appropriate inertial-range variables, such as the velocity-differences
$\delta v_\ell(x)=v(x+\ell)-v(x)$,
and depend only on the length-scale $\ell$ and the mean dissipation
$\varepsilon$ per mass.  However, much evidence
has accumulated subsequently that there is a nontrivial dependence of such PDFs
on the ratio $L/\ell$ \cite{interm}.
This is associated to an increase in the fluctuations, or ``intermittency,'' of
the inertial-range variable for
increasing ratio $L/\ell$, so that the first limit $L\rightarrow\infty$ appears
to lead to diverging moments.
It has also been questioned whether the second limit $\eta\rightarrow 0,$
physically associated to vanishing
viscosity $\nu\rightarrow 0,$ exists, or whether the limit is independent of
the particular form of dissipation,
such as hyperviscosity vs. ordinary viscosity. In a simple shell dynamical
model, it was found by Leveque and
She \cite{LS} that, while the limit $\nu\rightarrow 0$ appeared to exist, the
values of the scaling exponents
in the putative inertial range depended upon the degree of the hyperviscosity.
Subsequent numerical study of
Navier-Stokes turbulence \cite{CCS} found no similar dependence, although the
low Reynolds number of the
simulation makes such a result preliminary. It may be taken as the current
``conventional wisdom'' that the
limit $\eta\rightarrow 0$ exists and results in no ambiguity, while the limit
$L\rightarrow \infty$ does not
exist (at least for moments) and is a source of anomalous scaling, if not an
outright breakdown of universality.
However, the question which of the two limits exists, if either, and also the
uniqueness of the limits must be
regarded as open at the level of rigorous results. The issue frequently
generates heated debates (e.g. \cite{Wal}).

A simpler situation than the Navier-Stokes fluid for which many of the same
questions arise is the problem of the
turbulently convected passive scalar. The dynamics of the scalar field
$\theta(\br,t)$ is given by the linear equation
\be (\partial_t+\bv\bdot\grad_\br)\theta=\kappa\bigtriangleup_\br\theta+f
\lb{pseq} \ee
in which $\bv(\br,t)$ is the turbulent velocity and $f(\br,t)$ is a scalar
source at macroscale $L$. The closest analogy
exists for the so-called {\em inertial-convective range} of the passive scalar,
in which the advecting velocities exhibit
inertial-range scaling and the molecular diffusivity $\kappa$ is small, so that
the scalar dynamics itself is dominated by
convection. In other words, this range corresponds to the idealized limit of
infinite P\'{e}clet number, ${\rm Pe}
=UL/\kappa=\infty$ ($U$ is the typical velocity fluctuation at the macro
lengthscale). A dimensional theory for this
situation analogous to that of Kolmogorov was developed soon after his by
Obukhov and by Corrsin \cite{OC}. However,
similar evidence has been found to suggest that the Obukhov-Corrsin theory is
flawed in the same way as K41 \cite{Ant}:
while the existence of the $\kappa\rightarrow 0$ limit is not threatened by the
present data, the experiments suggest that
there is a nontrivial dependence upon $L$ due to intermittent cascade of the
scalar from the macroscale.

Recently, this ``conventional picture'' has obtained some support in a model of
the passive scalar, first considered by
Kraichnan in 1968 \cite{Kr68}. In this model, the (incompressible) velocity
$\bv$ and the source $f$ are both taken to be
Gaussian random fields delta-correlated in time. A regime mimicking the
inertial-convective range of true turbulence is
produced when the velocity covariance in space is taken to obey short-distance
scaling:
\begin{eqnarray}
\langle v_i(\br)v_j(\bz)\rangle & \sim & V_0\delta_{ij}-D\cdot
r^\zeta\cdot\left[\delta_{ij}^{\,} \right.\cr
\, & &  \,\,\,\,\,\,\,\,\,\,\,\,\,\,\,\,\,\,
\left.+{{\zeta}\over{d-1}}\left(\delta_{ij}-{{r_ir_j}\over{r^2}}\right)\right]
\lb{scaleq}
\end{eqnarray}
for $r\ll L.$  Kraichnan's ``rapid-change'' model is exactly soluble in the
sense that there is no closure problem, i.e.
the $N$th correlator of the scalar, $\Theta_N(\br_1,...,\br_N;t)$ obeys a
linear equation
\begin{eqnarray}
\partial_t\Theta_N & = & -\widehat{{\cal H}}_N\Theta_N+\sum_{{\rm
pairs}\,\,\,\,\{nm\}} F(\br_n,\br_m)\times \cr
\, & & \;\;\; \;\;\;\;\;\;\;\;\;\;\;\;\;\;\;\;\;\;\;\;\;
\Theta_{N-2}(...\widehat{\br_n}...\widehat{\br_m}...), \lb{closeq}
\end{eqnarray}
which involves only itself and lower order correlators. The linear operator
$\widehat{{\cal H}}_N$ is determined by the
velocity statistics (see Eq.(\ref{masmat}) below), while the inhomogeneous term
involves the source covariance $F$. The new
work  has involved either (i) a physically-motivated ansatz for dissipative
terms \cite{Kr94} or (ii) perturbative
exploration of limiting regimes: $\zeta\ll 1$ in \cite{GK} and
${{1}\over{d}}\ll 1$ in \cite{CF}. These studies all confirm
the ``conventional picture.'' Most recently \cite{CFL}, the
${{1}\over{d}}$-expansion has predicted an interesting
non-universality of the convective-range scaling exponents, through dependence
on the correlation time of the random
velocity field (when that time is sufficiently small). In addition, the above
theories all make detailed quantitative
predictions of convective range scaling exponents.  Unfortunately, the two
distinct approaches do not agree in detail
and neither has complete control of errors. Numerical simulations
\cite{Kr94,Kr96} have not yet been able to probe the
small $\zeta$ or large $d$ ranges where conflicting predictions occur. Thus,
the results, while exciting, are still
unconfirmed.

We give a very direct, nonperturbative proof of the correctness of the
``conventional picture'' of inertial-convective range
universality in the Kraichnan model. The full proofs are presented elsewhere
\cite{EX}, but the most significant details may
be easily explained here. The main step in our argument is to show that the
stationary solution of the model
Eq.(\ref{closeq}) is {\em unique} for the idealized $\kappa=0$ case, when a
certain regularity requirement is imposed. Our
proof is inspired by an analogy of the model with the {\em quantum many-body
problem}. In fact, the operator
$\widehat{{\cal H}}_N$ for $\kappa=0$ has the form of a quantum kinetic energy
operator in $Nd$ dimensions with a
position-dependent {\em mass matrix} $\bM(\bR)$, or
$\oH_N=(2\bM(\bR))^{-1}:\bP\bP.$ The mass matrix is defined in terms of
the velocity covariance as
\be \left[\bM(\bR)^{-1}\right]_{in,jm}=\langle v_i(\br_n)v_j(\br_m)\rangle,
\lb{masmat} \ee
where $\bR=(\br_1,...,\br_N)$ is the $N$-particle configuration point and
$i,j=1,...,d,\,\,n,m=1,...,N.$ $\bM(\bR)$ is
therefore a nonnegative matrix for every $\bR.$ Because of the
quantum-mechanical analogy it is very natural to look for
solutions among square-integrable functions ($\Theta_N\in \cL^2$). However, the
perturbation studies \cite{CF} suggest that
the decay at large $R$ is too slow for the solutions to be in $\cL^2.$ Since we
are only interested anyway in the
short-distance behavior, we confine the system to a bounded domain $\Omega$ in
$d$-dimensions, to remove this difficulty.
On the boundary of $\Omega$ we impose a fixed value $T_0$ of the scalar, as
would be appropriate in a real experiment with a
temperature field $T$ held fixed to $T_0$ by thermal contact at the channel
wall. Without loss of generality, we may take
the boundary condition $\theta_0=0$ (Dirichlet b.c.), by considering the
fluctuation field $\theta=T-T_0$ (as anticipated in
Eq.(\ref{pseq}).) We expect that our results will carry over also to more
artificial geometries, e.g. the periodic b.c.
used in numerical simulations \cite{Kr94}, although some parts of our proof
below depend upon the Dirichlet b.c.

There is a very simple formal argument which suggests the uniqueness of the
$\kappa=0$ solutions in $\cL^2.$ If we take $G_N$
to be the inhomogeneous term in Eq.(\ref{closeq}), then the stationary solution
of the equation should be given uniquely by
\be \Theta_N={\oH_N}^{-1}G_N. \lb{soln} \ee
To permit this formal inversion, two things are required. First, it must be the
case that $\oH_N$ has no zero-modes.
Secondly, it must also be shown that the range of $\oH_N$ is closed (Fredholm
condition). The insufficiency
of the first condition alone may be explained more physically by observing that
the inversion is still ill-defined if the
operator $\oH_N$ has no zero-modes but if $0$ is a point of accumulation of the
spectrum. In that case, the inverse
${\oH_N}^{-1}$ is not a bounded operator and the righthand side of
Eq.(\ref{soln}) need not exist. The Fredholm condition
is guaranteed, in particular, if the ``ground-state energy'' of $\oH_N$ is
strictly positive. Unfortunately, there is a real
danger that the condition is violated, because the mass matrix obviously has
infinite mass eigenvalues at certain points
$\bR.$ This certainly happens whenever two points coincide, $\br_n=\br_m$ for
$n\neq m,$ since then the velocity covariance
is degenerate. Such large masses lead to the possibility of small, positive
energies arbitrarily close to $0.$ Nevertheless,
we have shown \cite{EX} not only that zero-modes do not exist with Dirichlet
b.c. but also that the ``ground-state energy''
of $\oH_N$ is then positive. We may note, incidentally, that a similar
situation exists for periodic b.c. although (trivial)
zero-modes certainly occur there, the constants. An equation may be developed
for the {\em cumulant} part of the correlators,
$\Theta_N^c,$ of the form $\oH_N\Theta_N^c=G^c_N$, in which the inhomogeneous
term $G^c_N$ is defined in terms of
lower-order cumulants and is explicitly orthogonal to constants in the $\cL^2$
inner product. See Eq.(4.34) in \cite{GK-L}.
One therefore has a similar formal solution $\Theta_N^c={\oH_N}^{-1}G^c_N.$ As
above, this is indeed the unique
solution if (i) constants are the only zero-modes of $\oH_N$ and (ii) there is
a positive spectral gap above the 0 eigenvalue.
We prove below that (i) holds, but (ii) is still open for periodic b.c.

Establishing these facts requires a careful study of the mass matrix, carried
out in \cite{EX}. It is shown there
that zero eigenvalues of the inverse (Gramian) matrix $\bG(\bR)=\bM(\bR)^{-1}$
occur {\em only} when some points
$\br_n,\br_m$ coincide, or are ``fused.'' This result holds whenever the
velocity spectrum in Fourier space is everywhere
strictly positive. In general, the bad set of configurations $\bR$ where zero
eigenvalues occur has $K$ ``clusters'' of
coinciding particles with $N_k$ points in the $k$th cluster, $k=1,...,K.$ It is
shown in \cite{EX} that the precise number
of zero eigenvalues at each ``bad'' point is $\sum_{k=1}^K (N_k-1)d.$
Furthermore, a lower bound is derived for
the {\em least eigenvalue} $\nu(\bR)$ of $\bG(\bR)$ valid for all $\bR$, in
terms of the minimum distance
between any pair, $\rho(\bR)=\min_{n\neq m}r_{nm}$. (Note that
$r_{nm}=|\br_n-\br_m|$ is the distance between pairs
$\br_n,\br_m$. With periodic b.c. $r_{nm}$ should be taken to be the least
distance between all of their periodic images).
The estimate that is proved, for $0<\zeta<2$, is
\be \nu(\bR)\geq {\rm (const.)}[\rho(\bR)]^\zeta. \lb{lowbd} \ee
Establishing this bound is the crucial part of the proof. Its validity is
suggested by a simple perturbation theory
calculation in $[\rho(\bR)]^\zeta$ using the asymptotic short-distance formula
(\ref{scaleq}) for $\bG(\bR).$ What is
required is an argument that the first-order term in this expansion is
non-vanishing. The difficulty is that there is a
possible hierarchy of pair-separations $\{r_{nm}\}$ between the box size $L$
and the minimum length $\rho(\bR)$.
However, a careful examination of cases using the general first-order
perturbation formula and the short-distance
expression (\ref{scaleq}) show that the leading term is always the first-order
one. This argument yields Eq.(\ref{lowbd}).

It is not hard to show from the above results for $\bM(\bR)$ that constants are
the only zero-modes of $\oH_N.$
In fact, the stochastic representation holds that
$\langle\Theta_N,\oH_N\Theta_N\rangle={{1}\over{2}}\int d\bR \int d\bV
P_\bR(\bV)|(\bV\bdot\grad_\bR)\Theta_N(\bR)|^2$, with $P_\bR(\bV)$ the
multivariate Gaussian
with covariance $\bG(\bR)$ for the $Nd$-vector $\bV$. Because this probability
density is strictly positive
for every $\bR,\bV$ except for the ``bad'' points $\bR$ of zero measure, it
easily follows that any zero-mode
satisfies $\grad_\bR\Theta_N(\bR)=\bz$ almost everywhere. Hence, all zero-modes
are constants. In the case of
Dirichlet b.c. this rules out existence of any zero-modes at all.

The proof that the ``ground-state energy'' is strictly positive for Dirichlet
b.c. uses the lower bound Eq.(\ref{lowbd}),
giving
\be \langle\Theta_N,\oH_N\Theta_N\rangle\geq {\rm (const.)}\int
d\bR\,\,[\rho(\bR)]^\zeta|\grad_\bR\Theta_N(\bR)|^2
\lb{first} \ee
For any function $g(\bR)$ whose Laplacian $\bigtriangleup_\bR g$ is positive
and finite
except for a singularity manifold $\Gamma$ of codimension $\geq 2$, the
inequality holds that
\begin{eqnarray}
\, &  & \int d\bR\,\,|\bigtriangleup_\bR g(\bR)||\varphi(\bR)|^2 \cr
\, &  & \,\,\,\,\,\,\,
        \leq 4\int d\bR\,\,|\bigtriangleup_\bR g(\bR)|^{-1}|\grad_\bR
g(\bR)|^2|\grad_\bR\varphi(\bR)|^2,
\lb{Hardeq}
\end{eqnarray}
with $\varphi(\bR)$ any smooth function which $=0$ on the boundary
$\partial\Omega$. This is the result of an integration
by parts and the Cauchy-Schwartz inequality \cite{EX,Lew}. It may be applied to
$g(\bR)=[\rho(\bR)]^\zeta$ since
$\bigtriangleup_\bR g(\bR)=2\zeta(d-\gamma)[\rho(\bR)]^{-\gamma}$ with
$\gamma=2-\zeta$, which is positive as required
when $\zeta>0$ and $d\geq 2$. The Laplacian of $g(\bR)$ has singularities
precisely on the ``bad'' set of points $\bR$
with codimension $d\geq 2$ where mass eigenvalues become infinite. Thus, taking
$g(\bR)=[\rho(\bR)]^\zeta$ and
$\varphi(\bR)=\Theta_N(\bR)$ one obtains finally, if $\zeta>0$,
\begin{eqnarray}
\langle\Theta_N,\oH_N\Theta_N\rangle
          & \geq  & {\rm (const.)}{{(d-\gamma)^2}\over{2}}\times  \cr
\, & & \,\,\,\,\,\,\,\,\,\,\,\,\,\,\,\,\,\,
\int d\bR\,\,[\rho(\bR)]^{-\gamma}|\Theta_N(\bR)|^2
\lb{second}
\end{eqnarray}
for a set of $\Theta_N(\bR)$ dense in $\cL^2$. Because $\rho(\bR)\leq L$
everywhere, this implies that for $\gamma\geq 0$
\be \langle\Theta_N,\oH_N\Theta_N\rangle
          \geq {\rm (const.)}{{(d-\gamma)^2}\over{2}}L^{-\gamma}\|\Theta_N\|^2.
\lb{third} \ee
The last equation implies immediately a lower bound on the spectrum $\sim {\rm
(const.)}L^{-\gamma}$ whenever
$0<\zeta <  2$ and $d\geq 2.$ In fact, by a criterion of Friedrichs, it can be
shown from the two estimates
Eqs.(\ref{first}) and (\ref{second}) that the entire spectrum of $\oH_N$ is
discrete. See \cite{Lew}.
Thus, the solution at $\kappa=0$ is given uniquely in $\cL^2$ by
Eq.(\ref{soln}) for Dirichlet b.c. Unfortunately,
this proof does not work for periodic b.c. because the function $g(\bR)$ has
additional singularities on the
codimension-1 set of points where $r_{nm}=r_{n^*m}$, with $\br_{n^*}$ a
periodic image of $\br_n$. These singularities
contribute a surface term in the integration by parts argument which vitiates
the derivation of the inequality
Eq.(\ref{Hardeq}). Despite the failure of this simple proof, we expect that
there is still a gap in the
spectrum above $0$ for periodic b.c. and all our results remain true.

The proof of a unique solution in $\cL^2$ for $\kappa>0$ can be given along the
same lines but is much easier, since there
is then obviously a spectral gap of size $\sim {{\kappa}\over{L^2}}$. This
solution formally corresponds to a state
of zero Prandtl number, ${\rm Pr}=\nu/\kappa=0.$ At high-wavenumbers it
exhibits an infinitely long ``inertial-diffusive
range'' whose scalar spectrum decays with a large power. Such a range was
proposed by Batchelor et al. in \cite{BHT}
and recently studied in the ``rapid-change'' model \cite{FW}. This range
follows after a lower wavenumber inertial-convective
range a l\'{a} Obukhov-Corrsin.  Subsequently, one may consider the infinite
P\'{e}clet number limit, in which the high
wavenumber end of the inertial-convective range will move off to plus infinity.
It is not hard to show that limits of the
positive-$\kappa$ solutions exist along subsequences for $\kappa\rightarrow 0$
and are $\cL^2$ solutions of the $\kappa=0$
equations \cite{EX}. Using the uniqueness of the $\kappa=0$ solutions, it then
follows that the $\kappa=0$ solution is the
limit of the positive-$\kappa$ solutions. It therefore corresponds to an
idealized inertial-convective range of infinite
extent. The same argument applies to a broad class of dissipation terms besides
the standard diffusion, including
``hyperdiffusions'' $\kappa_p(-\bigtriangleup_\br)^p\theta.$ The conclusion is
obtained that, for all these dissipation
mechanisms, the limits exist in $\cL^2$ and coincide with the unique $\kappa=0$
solution (and therefore with each other).
Universality is thus established.

There is a simple physical interpretation of our results which is worth
emphasizing. Even in the limit as
molecular diffusivity vanishes, $\kappa\rightarrow 0$, there is an effective
diffusivity, or so-called
{\em eddy-diffusivity}, acting on the mean scalar statistics which is generated
through advection by the random velocity.
For $N$-point statistics with $N\geq 2$ the effective diffusivity is
scale-dependent, with  $\kappa_{{\rm eddy}}(r)\sim
D\cdot r^{\zeta}$ at length-scale $r$. This was first observed for 2-point
statistics (pairs of Lagrangian tracers) by
Richardson \cite{Rich}. He postulated a diffusion equation of the same form as
Eq.(\ref{closeq}) with an
eddy-diffusivity tensor $K_{ij}(r)$ scaling as Eq.(\ref{scaleq}) for
$\zeta={{4}\over{3}}$. The matrix $\bG(\bR)$
is the $N$-point generalization of the Richardson tensor and the
Eq.(\ref{lowbd}) for its lowest eigenvalue is just
an expression of the expected scaling for $\kappa_{{\rm eddy}}(r)$. Note that
the $L$-dependence of the spectral
gap for $\kappa=0$ is a direct consequence, since it should plausibly scale as
${{\kappa_{{\rm eddy}}(L)}\over{L^2}}\sim
L^{-\gamma}$. It is important to observe that these eddy-diffusivity effects
are only demonstrated for statistical
correlations and not at the level of individual realizations of the scalar
field. This explains the paradoxical-looking
fact that as $\kappa\rightarrow 0$ the b.c. are satisfied by the statistical
correlations, solving Eq.(\ref{closeq}),
but not necessarily by the individual scalar fields, with dynamics
Eq.(\ref{pseq}). Even if the scalar source $f$ is removed,
it may not be possible to maintain the Dirichlet b.c. on $\theta(\br,t)$ if
$\kappa=0$. For such ideal dynamics the
scalar field (when smooth) is simply transported unchanged along Lagrangian
characteristics. Any point $\br$ in the
interior of $\Omega$ with initial $\theta_0(\br)\neq 0$ which subsequently
flows into the boundary $\partial\Omega$ will
violate the b.c. at that point. For $\kappa>0$ the b.c. on $\theta$ are
maintained by the action of the molecular diffusion
on the strong gradients of scalar concentration formed by such advection to the
boundary. This results in a thin ``diffusive
boundary layer'', with thickness vanishing as $\kappa\rightarrow 0$, across
which the scalar intensity drops rapidly to
zero. For our model, the b.c. for statistical correlations are maintained in
the same way by the eddy-diffusivity, whose
action creates a ``turbulent boundary layer'' across which $N$-point functions
drop to zero.

The conclusions obtained here hopefully extend to the true passive scalar and,
even more optimistically, to the
Navier-Stokes fluid velocity. It is, at least, satisfying to have one model in
which the ``conventional wisdom''
for these turbulent systems may be verified without question. Our analysis has
some further interest
in terms of the perturbative studies \cite{GK,CF}. Those works obtained
nontrivial zero-modes of
local approximations to $\oH_N$ by means of a Rayleigh-Schr\"{o}dinger
perturbation theory. This does not
contradict our result that all zero-modes are constants, because we consider
the {\em global} operator.
Our result just states that ambiguity in the perturbative treatments are
removed by matching to the
macro-scale $L$. In fact, our results confirm one of the basic assumptions used
in the perturbative treatments:
explicit matching to the dissipation scale may be replaced by a requirement of
short-distance regularity.
Even the weak requirement of local square-integrability is sufficient to select
the solution properly matched
to the dissipation range.

{\it Acknowledgements:} We wish to thank R. H. Kraichnan for extremely helpful
discussions of this problem
and U. Frisch for comments on the paper.  J. X. is partially supported by NSF
grant DMS-9302830.

\end{document}